\documentclass[12pt]{article}
\usepackage{latexsym}
\usepackage{epsfig}

\newcommand{\del}{\partial}
\newcommand{\beq}{\begin{eqnarray}}
\newcommand{\eeq}{\end{eqnarray}}
\newcommand{\be}{\begin{eqnarray*}}
\newcommand{\ee}{\end{eqnarray*}}
\newcommand{\bk}{{\bf k}}

\newcommand{\bx}{{\bf x}}

\newcommand{\ra}{\rightarrow}
\newcommand{\e}{\epsilon}
\newcommand{\ve}{\varepsilon}
\newcommand{\nn}{\nonumber}

\newcommand{\ex}[1]{\langle\,#1\,\rangle}

\def\square{\vcenter{\vbox{\hrule height.4pt
          \hbox{\vrule width.4pt height6pt
          \kern6pt\vrule width.4pt}\hrule height.4pt}}}

\begin{document}

\centerline{\Large\bf {Problems with the Casimir Vacuum Energy}}
\vskip 5mm
\centerline{Finn Ravndal\footnote{Extended version of contributed talk at {\it Vacuum Energy and the
       Cosmological Constant}, NORDITA, Copenhagen, August 24 - 26, 2000.}}
\vskip 5mm
\centerline{\it Institute of Physics, University of Oslo, N-0316 Oslo, Norway.}

\begin{abstract}

A critical look is taken at the calculation of the Casimir effect. The boundary conditions
play an important role and should be imposed in a physical way. An acceptable result for the
vacuum energy is only obtained when different regularization schemes yield the same result.
Radiative corrections to the Casimir force between two parallel plates due to electromagnetic 
vacuum fluctuations have been obtained both in full QED and in a low-energy, effective field
theory with conflicting results. Including surface terms in the effective Lagrangian, the QED
result can be understood. Finally, a problem with the electromagnetic vacuum energy on $S^3$ is
pointed out and a solution suggested. The structure of higher order corrections to this result
is discussed.

\end{abstract}

\section*{Introduction}

In modern physics one has two main mechanisms which contribute to the energy of the vacuum. One  
is due to the condensates of various fields or operators when certain symmetries are spontaneously
broken. This is essentially a classical effect. The other one, which we will consider here, is the
zero-point energy. It is a direct consequence of quantum mechanics. In the canonical formulation
of Heisenberg and Schr\"odinger it arises from a certain ambiguity in the ordering of field 
operators while in the functional formulation of Feynman it arises from a corresponding ambiguity
in defining the path integral in a continuous spacetime. Since the birth of modern quantum mechanics
a central question has been if these ambiguities can have physical consequences in the sense of being
measureable. 

One of the most remarkable manifestations is the attractive force between two parallel 
metallic plates in vacuum induced by the fluctuations in the electromagnetic field. It was first 
calculated by Casimir\cite{theory} and has now been experimentally confirmed in two independent 
experiments\cite{exp}. Higher precision in such experiments are within sight and it will now become
necessary to take a closer look at the theory behind this important phenomenon and also the
experiments. Here we will not discuss the latter except to point out that they are not done
in the original geometry of two parallel plates for which the force has been calculated, but with a 
spherical surface above a metal plane for which the effects of vacuum fluctuations are not known.
Instead one relates this unknown situation to the first one by the use of a classical proximity
theorem. This should be analyzed more carefully in a quantum context.

The energy of a free quantum field with mass $m$ in a certain state characterized by the 
occupations numbers $\{n_\bk\}$ is
\be
    E = \sum_\bk\,(n_\bk + {1\over 2})\,\omega_\bk
\ee
where $\omega_\bk = (\bk^2 + m^2)^{1/2}$ is the energy of a particle with momentum $\bk$. In the
vacuum all the modes are empty and one gets only a contribution from the last zero-point term
\beq
    E_0 = {1\over 2}\sum_\bk\,\omega_\bk                         \label{E_0}
\eeq
This is a divergent constant which is usually just neglected or removed by other more theoretical 
constructions invoking the argument that we are only interested in energy differences. In the
absence of gravity and in homogeneous spaces this is justified, but in general it is not.

Replacing the sum over discrete modes in (\ref{E_0}) in an open space with the corresponding 
integral, we see that the divergence
is quartic. For very large momenta the particle mass can be ignored. Imposing an exponential cutoff
$\Lambda$, we thus have for the resulting vacum energy density ${\cal E} = E_0/V$ a finite result
\beq
    {\cal E} = \int\!{d^3k\over (2\pi)^3}\,\omega\, e^{-k/\Lambda} 
             \approx {3\Lambda^4\over 2\pi^2}                           \label{CC}
\eeq
It has the same value everywhere and is therefore a contribution to the cosmological constant. The
recent supernova observations can be explaned by such a vacuum energy with a value ${\cal E}_{observ} 
= \mu^4$ where the mass $\mu \approx 10^{-3}$\,eV. From the point of view of effective field theory, 
the cutoff $\Lambda$ represents the energy scale where the applicapability of the above theory ceases. 
Taking it to be the Planck mass $M \approx 10^{19}$\,GeV where the classical spacetime picture
breaks down, we thus find the ratio of the theoretical to the observed cosmological constant to be
\beq
     {{\cal E}_{theory}\over{\cal E}_{observ}} \approx \left({\Lambda\over\mu}\right)^4 
                                          \approx 10^{124}
\eeq
This is usually called the worst theoretical prediction ever in modern physics. But it is really not 
that bad. A naive calculation of the self-energy of a particle will also give a huge result depending
on a cutoff as long as the physics at short scales is not handled consistently. The same applies
here. The above cutoff is just a reflection of our ignorance of physics at the 
shortest scales. When that is understood in terms of strings, branes or something more fundamental,
this apparent failure will be replaced by predictive power.

\section*{Casimir vacuum energy}
The vacuum fluctuations  between two parallel plates shown in the Fig. 1 normal to the $z$-axis and 
separated by the distance $L$ with the left plate at $z = 0$, gives rise to an attractive Casimir 
force between the plates. It is most simply described for a massless scalar field $\phi(\bx,t)$ with 
the wave equation $\del^2\phi = 0$.
\begin{figure}[htb]
  \begin{center}
    \epsfig{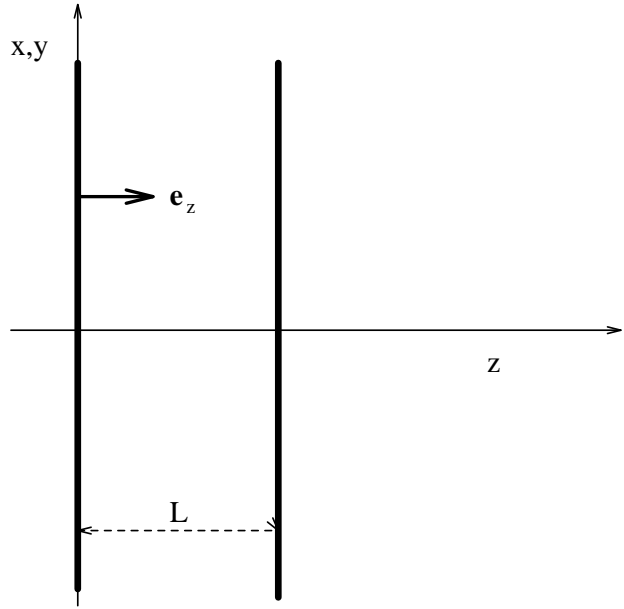}
  \end{center}
  \vspace{-4mm}            
  \caption{\footnotesize The fluctuating quantum field is confined beween two parallel plates.} 
  \label{fig}
\end{figure}
A mode with energy $\omega$, i.e. $\phi({\bf x},t) = u({\bf x})\exp(i\omega t)$, is therefore 
given by the solution of the Helmholtz equation $(\nabla^2 + \omega^2)u({\bf x}) = 0$. In order to
quantize the field between the plates, we need to know the boundary conditions it satisfies at the
plates. They must be arrived at from some physical or mathematical requirement imposed on the system.
If we choose to use Dirichlet boundary conditions, i.e.  $\phi({\bf x},t)|_{z=0} = 
\phi({\bf x},t)|_{z=L} = 0$, the corresonding eigenmodes are
\beq
     u_{n{\bf k}_T}^D({\bf x}) = \sin{k_zz}\,e^{i{\bf k}_T\cdot{\bf x}_T}       \label{Diri}
\eeq
where $\bx_T = (x,y)$. The component of the wavenumber along the $z$-axis is now quantized with the
values $k_z = (\pi/L)n$ where $n = 1,2,3,\cdots$. Each such mode thus has the energy 
$\omega_{n{\bf k}_T} = ({\bf k}_T^2 + (\kappa\,n)^2)^{1/2}$ with $\kappa = \pi/L$. 

The total vacuum energy (\ref{E_0}) between the plates is now
\beq
   E_0^D = A \int\!{d^2k_T\over(2\pi)^2}\sum_{n=1}^\infty
           {1\over 2}\;\omega_{n{\bf k}_T}		               \label{casa}
\eeq                        
where $A$ is the cross-sectional area of one plate. Replacing $k_T$ with $\omega$ as a new integration
variable and using that $k_T dk_T = \omega d\omega$, we have
\be
   E_0^D = {A\over 4\pi}\sum_{n=1}^\infty \int_{\kappa n}^\infty\!d\omega\,\omega^2
\ee
In order to regularize the divergences both in the integral and the sum, we impose an exponential
cutoff $\Lambda = 1/\e$. Then it follows
\be
      E_0^D & = &  {A\over 4\pi}\sum_{n=1}^\infty \int_{\kappa n}^\infty\!
           d\omega\,\omega^2  e^{-\ve\omega} 
      =    {A\over 4\pi}\sum_{n=1}^\infty {\partial^2\over\partial\ve^2}
      \int_{\kappa n}^\infty\!d\omega\,e^{-\ve\omega} \nn \\	     
      &=&  {A\over 4\pi} {\partial^2\over\partial\ve^2}
      {1\over \ve}\sum_{n=1}^\infty e^{-\ve n\kappa}
      =	{A\over 4\pi} {\partial^2\over\partial\ve^2}{1\over \ve}
                         {1\over e^{\kappa\ve} - 1}    	                        \label{cas.7}
\ee
We can now isolate a pure vacuum divergence in the limit $\Lambda \ra\infty$ by using the 
Bernoulli expansion
\be
     {x\over e^x - 1} =\sum_{n=0}^\infty {B_n\over n!}\,x^n  = 1 - {x\over 2} + {x^2\over 12}
                      - {x^4\over 720} + \cdots                                
\ee
when $x\ra 0$. It gives for the energy density  ${\cal E} = E_0/AL$ with pure Dirichlet boundary 
conditions the result
\beq
     {\cal E}^D = {3\Lambda^4\over 2\pi^2} - {\Lambda^3\over 4\pi L}
                - {\pi^2 \over 1440 L^4}                                          \label{cas.10}
\eeq
It is seen that the first term is the divergent vacuum energy (\ref{CC}) in empty space which is 
recovered in the limit $L \ra \infty$ when the plates are removed away from each other. While the 
second term contains the ultraviolet cutoff, the last term is finite and is said to be the Casimir 
energy density between the plates. 

The troublesome middle term is usually not seen or argued away by introduction of corresponding 
counterterms. For instance, if we use dimensional regularization where the transverse integration in
(\ref{casa}) is done in $d$ dimensions which are taken to be so small that the integral is convergent,
one obtains
\be
      E_0^D =  {A\over 2}\sum_{n=1}^\infty {\kappa^{d + 1}\,\Gamma(-{1+d\over 2})\over
               (4\pi)^{\,d/2}\,\Gamma(-{1\over 2})}
\ee
Now analytically continuing back to the physical situation where $d=2$, we find
\beq
       E_0^D = - {A\over 12\pi} \left({\pi\over L}\right)^3\sum_{n=1}^\infty n^3 
             = - {A \pi^2\over 1440 L^3}                                        \label{zeta}
\eeq
when the divergent sum is regularized by  Riemann's zeta-function to be $\zeta(-3) = 1/120$. We have 
thus recovered the finite Casimir energy found as the last term in (\ref{cas.10}). The divergent,
$L$-independent term is absent which we consider to be an attractive aspect of this 
regularization scheme since it corresponds to subtracting a universal, constant part. But the absence
of also the middle term is a peculiarity of this particular regularization scheme and most likely 
does not reflect any underlying physics. A believeable, physical result should be obtained with
the same value independent of the regularization scheme used.

Recently Hagen has raised some criticism against the regularization methods used in the calculation
of the Casimir energy\cite{hagen}. Although most of his arguments have now been countered\cite{brevik},
one should still be careful in making too rapid mathematical conclusions about vacuum fluctuations 
whose physical existence still requires much experimental work. As an example, we can return to the
scalar field considered above for which we imposed the Dirichlet boundary conditions. But there is
nothing sacred about them. In fact, we could instead choose to work with Neumann conditions so that
the normal derivative of the field vanishes at the plates, i.e. $\partial_z\phi({\bf x},t)|_{z=0} =
\partial_z\phi({\bf x},t)|_{z=0} = 0$. The eigenmodes are then
\beq
     u_{n{\bf k}_T}^N({\bf x}) = \cos{k_zz}\,e^{i{\bf k}_T\cdot{\bf x}_T}         \label{Neum}
\eeq
where now $n = 0,1,2,\cdots$ and the mode energies are obviously the same. The total energy is 
therefore again given by (\ref{casa}) where
now the sum starts at $n=0$. With dimensional and zeta-function regularization one then gets the
same result (\ref{zeta}) as before while with an exponential cutoff we obtain
\beq
     {\cal E}^N = {3\Lambda^4\over 2\pi^2} + {\Lambda^3\over 4\pi L}
                - {\pi^2 \over 1440 L^4}                                          \label{cas.20}
\eeq
instead of (\ref{cas.10}). The troublesome term thus appears with opposite sign. One can show 
that the fluctuation in the field now diverges near the plates with opposite sign compared with the 
previous Dirichlet case\cite{Dag}. Also the energy density between the plates is different and
should be calculated including the conformal term in the scalar Lagrangian describing a massless
field.

This suggests a new approach to the Casimir energy for scalar fields. One should always consider the 
contribution from actually two fields, one with Dirichlet and one with Neumann boundary conditions. 
Combined, we then have the physically meaningful result
\beq
     {\cal E}^{D+N} = {3\Lambda^4\over \pi^2}  - {\pi^2 \over 720 L^4}            \label{cas.30}
\eeq
which now agrees with what one also obtains with dimensional regularization. This
point of view is also consistent with a more detailed investigation by Symanzik\cite{Kurt}. 
Or alternatively, we must have two scalar fields as members of a larger multiplet where each satisfies
one of the above two boundary conditions. This is the case in the Wess-Zumino supersymmetric theory 
where they appear together with a massless fermion field\cite{japs}. The total vacuum energy from all 
three fields is then finite and actually zero as it should be from supersymmetry.

In order to calculate the Casimir energy for the electromagnetic field $({\bf E},{\bf B})$ one must 
first find the modes satisfied by the wave equation with the standard metallic boundary conditions
${\bf n}\wedge{\bf E} = {\bf n}\cdot{\bf B} = 0$  where the unit vector ${\bf n}$ is normal to the 
plates in Fig. 1. The modes can then be classified in terms of electromagnetic multipole 
fields\cite{LR}. 
In fact, the transverse electric multipoles TE with longitudinal quantum numbers
$n = 1,2,3,\cdots$ correspond to the Dirichlet modes (\ref{Diri}) while the transverse magnetic 
multipoles TM with $n = 0,1,2,\cdots$ correspond to the Neumann modes (\ref{Neum}). Without the
special TEM Neumann modes with $n=0$, one would not have obtained a finite result for the vacuum
energy. The same applies to the field fluctuations $\ex{{\bf E}^2}$ and $\ex{{\bf B}^2}$ which
actually varies with position between the plates, their sum being constant.

\section*{Radiative corrections}

The above Casimir energies have been calculated with free fields. Interactions have been considered
in the scalar case by Symanzik\cite{Kurt}, while radiative corrections to the total vacuum energy
for the electromagnetic field between parallel plates have been calculated by Bordag 
{\it et al.} within full QED\cite{Bordag}. The photon field satisfies standard boundary conditions
while the electron field is assumed not be affected by the presence of the plates. It is a
technically very demanding calculation which more recently has been confirmed in a somewhat
simpler approach\cite{BS}. They find that the Casimir force is modified by a small term going like
$\alpha/mL^5$ where $m$ is the electron mass. It results from contributions at a very short distance
scale of the order $1/m$.

In the meantime the same problem had been considered from the point of view of  effective field 
theory\cite{KR}. For energies well below the electron mass, one can integrate out the electron field 
and thus obtain the derivative expansion of the effective Lagrangian,
\beq
     {\cal L} =  - {1\over 4}F_{\mu\nu}^2 
              -  {\alpha\over 60\pi m^2}\,\del^\lambda F^{\mu\nu}\del_\lambda F_{\mu\nu} 
              +  {\alpha^2\over 90m^4}\left[(F_{\mu\nu}F^{\mu\nu})^2 
              +  {7\over 4}(F_{\mu\nu}\tilde{F}^{\mu\nu})^2\right] +\ldots          \label{LEH}
\eeq
where $\tilde{F}_{\mu\nu} = {1\over 2}\epsilon_{\mu\nu\rho\sigma}F^{\rho\sigma}$ is the dual 
field strength. In open space the middle term, which gives the Uehling effect of vacuum polarization,
can be removed by a field redefinition. One is thus left with the last term which is the 
Euler-Heisenberg interaction. It allows a simple calculation of the radiative correction to the 
Stefan-Boltzmann energy density for black-body radiation which is seen by inspection of the Lagrangian 
to be of the order $(\alpha^2/m^4)\,T^8$ where $T$ is the temperature\cite{KR}.
  
Since the calculation of finite-temperature effects corresponds to a Casimir calculation with
parallel plates normal to the imaginary time axis with periodic boundary conditions, it was
therefore natural to assume that the same Euler-Heisenberg effective Lagrangian also would give
the leading order corrections to the Casimir effect with real plates normal to the $z$-axis\cite{KR}.
One thus finds a result for the correction to the Casimir force which goes like $\alpha^2/m^4L^8$. 
Needless to say, it is in strong disagreement with the result obtained by Bordag 
{\it et al.}\cite{Bordag}. Such a discrepancy would be one of the very few known cases where there is a
disagreement between results derived in an effective theory and those derived from the underlying 
and more fundamental theory.

There is a way out of the above conflict. The effective Lagrangian must in general be constructed
from the degrees of freedom involved at the scale we consider and made to satisfy the symmetries
present in the problem. In our case it must therefore involve the gauge invariant field tensor
$F_{\mu\nu}$ and derivatives thereof. While the Lagrangian (\ref{LEH}) is Lorentz-invariant, this is
not really a symmetry present in the Casimir problem since translational invariance along the
$z$-axis is destroyed by the presence of the plates. As recently suggested by Thomassen\cite{Jan},
this can be compensated by the introduction of the 4-vector $n^\mu = (0,{\bf n})$ which in the 
effective theory is to be considered as a remnant of the boundary conditions in the full theory. 
With this vector in addition to the gradient operator $\del_\lambda$, we can now construct new, 
effective interactions like
\beq
      \Delta{\cal L} = C_0\, n^\lambda F^{\mu\nu} n_\lambda F_{\mu\nu} 
                     + {C_1\over m} n^\lambda F^{\mu\nu} \del_\lambda F_{\mu\nu} + \ldots
\eeq
where $C_0$ and $C_1$ are dimensionless coupling constants. The first term will give a correction to
the leading order Casimir force varying like $1/L^4$, while the second term
will give a result varying with the plate separation as $1/mL^5$ in agreement with the result of
Bordag {\it et al.}\cite{Bordag}. The unknown couplng constants appearing in this effective description
can be obtained from matching to suitable quantities calculated in full QED. When these are 
established, one can then investigate how the energy and pressure is distributed between and outside
the plates\cite{RT}.

\section*{Vacuum energy on $S^3$}

A non-zero vacuum energy can not only be caused by the presence of boundaries which disrupt the
free modes, but also by a non-trivial topology of spacetime. One of the simplest examples which
was first considered by Ford\cite{Ford}, is the massless scalar field on $S^3$. The conformally
invariant Klein-Gordon equation is
\beq
      (\Box + {1\over 6}R)\phi(x) = 0
\eeq
where $\Box$ is the covariant d'Alembertian and $R$ is the scalar curvature $R = R^\mu_{\;\,\mu}$. 
For the static spacetime on $S^3$ with radius $a$, the Ricci tensor is $R_{ij} = 2g_{ij}/a^2$ so that 
$R = 6/a^2$. The isometry group of this homogeneous space is $SO(4)$ and thus one can label the modes 
by the quantum numbers $n = 1,2,3,\ldots$ and $\ell = 0,1,\ldots,n-1$ as for the energy levels 
in the hydrogen atom. Each mode with energy $\omega = n/a$ thus has a degeneracy of $n^2$ and the 
total vacuum energy is
\beq
    E_0 = {1\over 2a}\sum_{n=1}^\infty n^3
\eeq
Using zeta-function regularization as in (\ref{zeta}), we therefore find $E_0 = 1/240 a$ in full
agreement with Ford who used an exponential cutoff.

For the electromagnetic Casimir energy on $S^3$ one would intuitively expect twice this energy
because of the two spin degrees of freedom and conformal invariance. However, Ford finds the
mode energies $\omega = n/a$ now with quantum numbers $n = 2,3,\ldots$ and degeneracy $2(n^2 -1)$.
Obviously, this will not give the expected vacuum energy although it is finite with zeta-function 
regularization. On the other hand, it is not finite using an exponential cutoff and thus
according to our previous discussions, there may be something wrong in the present description of the
problem. An obvious remedy is to locate some missing $\ell = 0$ modes in the solutions of the 
vector wave equation on $S^3$ so that the range of the quantum number $n$ and the degeneracy come 
out as in the scalar case\cite{Svend}. These would then play the same crucial r\^ole here as the TEM
mode with $n=0$ does for the electromagnetic Casimir energy between parallel plates.

When this problem is settled, it would be interesting to consider the radiative corrections to the
vacuum energy. The structure of the result can be analyzed from the point of view of effective 
field theory. From the requirements of general covariance and gauge invariance, the effective
Lagrangian must then depend on the field and curvature tensors. The derivative expansion will 
be in terms of the covariant gradient $\nabla_\mu$. In this way one can construct a leading order 
effective theory of the form
\beq
     {\cal L} &=&  - {1\over 4}F_{\mu\nu}F^{\mu\nu} \nn  \\
              &+&  {\alpha\over m^2}\left(C_1\nabla^\lambda F^{\mu\nu}\nabla_\lambda F_{\mu\nu}
     + C_2F^\mu_{\;\;\sigma}R^\sigma_{\;\;\rho}F^\rho_{\;\;\mu} 
     + C_3 F^{\mu\nu}F^{\rho\sigma}R_{\mu\nu\rho\sigma} + \cdots\right)            \label{LEFT}
\eeq
A common factor of $\alpha/m^2$ is taken out from the coupling constants because these contributions
will arise from Feynman diagrams involving two electron propagators. Together they will thus give a 
correction to the leading order Casimir energy which will vary as $\alpha/m^2a^6$. The precice value 
of the coefficient can be obtained by determining the unknown, effective coupling constants $C_1,
C_2,\ldots$ by matching to the underlying theory which is QED on this compact space.

\section*{Conclusion}

With the recent impressive experimental investigations of the Casimir effect, we will also need
a better understanding on the theoretical side. Soon we will have measurements of such an accuracy that
thermal effects must also be included. The underlying quantum vacuum fluctuations are
known to become very large near confining boundaries. Similar divergences will be found in most
components of the energy-stress tensor, in particular when interactions are included. Most of the
theoretical work so far has been concentrated mainly on the total Casimir energy. Much more 
information and physics to understand is hidden in the energy density and how it varies with position.
For this purpose effective field theory seems to be a most promising theoretical approach.


\begin{thebibliography}{99}

\bibitem{theory} H.B.G. Casimir, {\it Proc. K. Ned. Akad. Wet.} {\bf 51}, 793 (1948).

\bibitem{exp} S.K. Lamoreaux, {\it Phys. Rev. Lett.} {\bf 78}, 5 (1997); U. Mohideen and A. Roy,
              {\it ibid.} {\bf 81}, 4549 (1998).

\bibitem{hagen} C.R. Hagen, {\it Phys. Rev.} {\bf D61}, 065005 (2000); 
                {\it quant-ph/0003108, hep-th/0004079.}

\bibitem{brevik} I. Brevik, B. Jensen and K.A. Milton, {\it hep-th/0004041.}

\bibitem{Dag} D. Tollefsen, {\it Cand. Scient. thesis}, University of Oslo, 1988.

\bibitem{Kurt} K. Symanzik, {\it Nucl. Phys.} {\bf B190}, 1 (1981).

\bibitem{japs} Y. Igarashi, {\it Phys. Rev.} {\bf D30}, 1812 (1984);  Y. Igarashi and T. Nonoyama,
               {\it Prog. Theor. Phys.} {\bf 77}, 427 (1987).

\bibitem{LR} C.A. L\"utken and F. Ravndal, {\it Phys. Rev.} {\bf A31}, 2082 (1985).

\bibitem{Bordag} M. Bordag, D. Robaschik and E. Wieczorek, {\it Ann. Phys.} {\bf 165}, 192 (1985);
                 D. Robaschik, K. Scharnhorst and E. Wieczorek, {\it ibid.} {\bf 174}, 401 (1987).

\bibitem{BS} M. Bordag and K. Scharnhorst, {\it Phys. Rev. Lett.} {\bf 81}, 3815 (1998).

\bibitem{KR} X. Kong and F. Ravndal, {\it Phys. Rev. Lett.} {\bf 79}, 545 (1997).

\bibitem{Jan} J. Thomassen, {\it private communication.}

\bibitem{RT} F. Ravndal and J. Thomassen, {\it in preparation}.

\bibitem{Ford} L.H. Ford, {\it Phys. Rev.} {\bf 11}, 3370 (1975).

\bibitem{Svend} S. Hjelmeland and F. Ravndal,  {\it in progress}.

\end{thebibliography}
\end{document}